\documentclass[sigconf]{acmart}
\AtBeginDocument{%
  }

\setcopyright{acmlicensed}
\copyrightyear{2018}
\acmYear{2018}
\acmDOI{XXXXXXX.XXXXXXX}
\acmISBN{978-1-4503-XXXX-X/2018/06}

\usepackage[most]{tcolorbox}

\newcounter{findingcount}

\newtcolorbox[auto counter, number within=section]{findingbox}[1][]{
    enhanced,
    colback=gray!8,
    colframe=gray!50,
    arc=0pt,
    outer arc=0pt,
    boxrule=0.4pt,
    left=4pt, right=4pt, 
    top=2pt, bottom=2pt,
    before skip=4pt, 
    after skip=4pt,
    fontupper=\rmfamily,
}

\newtcolorbox{summarybox}{
    enhanced,
    colback=blue!5,          % Subtle blue tint to distinguish from findings
    colframe=blue!40!gray,   % Muted blue-gray frame
    arc=0pt,
    outer arc=0pt,
    boxrule=0.6pt,           % Slightly thicker than findingbox to denote a "summary"
    left=6pt, right=6pt, 
    top=4pt, bottom=4pt,
    before skip=6pt, 
    after skip=6pt,
    fontupper=\rmfamily,
}
\begin{document}

\title{Towards Explainable Stakeholder-Aware Requirements Prioritisation in Aged-Care Digital Health }

\author{Yuqing Xiao}
\email{yuqing.xiao@monash.edu}
\orcid{0000-0003-2922-3614}
\affiliation{%
  \institution{Monash University}
  \city{Melbourne}
  \state{Victoria}
  \country{Australia}
}

\author{John Grundy}
\email{john.grundy@monash.edu}
\orcid{0000-0003-4928-7076}
\affiliation{%
  \institution{Monash University}
  \city{Melbourne}
  \state{Victoria}
  \country{Australia}}
\author{Anuradha Madugalla}
\orcid{0000-0002-3813-8254}
\affiliation{%
  \institution{Deakin University}
  \city{Melbourne}
  \state{Victoria}
  \country{Australia}}
\author{Elizabeth Manias}
\email{elizabeth.manias@monash.edu}
\orcid{0000-0002-3747-0087}
\affiliation{%
\institution{Monash University}
\city{Melbourne}
\state{Victoria}
\country{Australia}}

\renewcommand{\shortauthors}{Xiao et al.}

\begin{abstract}
Requirements engineering (RE) for aged-care digital health must account for human aspects, because requirement priorities are shaped not only by technical functionality but also by stakeholders’ health conditions, socioeconomics, and lived experience. Knowing which human aspects matter most, and for whom, is critical for inclusive and evidence-based requirements prioritisation. Yet in practice, while some studies have examined human aspects in RE, they have largely relied on expert judgement or model-driven analysis rather than large-scale user studies with meaningful human-in-the-loop validation to determine which aspects matter most and why. To address this gap, we conducted an explanatory mixed-methods study with 249 participants (103 older adults, 105 developers, and 41 caregivers). We first applied SHAP-based explainable machine learning to identify the human aspects most strongly associated with requirement priorities across eight aged-care digital health themes, and then conducted 12 semi-structured interviews to validate and interpret the quantitative patterns. The results identify the key human aspects shaping requirement priorities, reveal their directional effects, and expose substantial misalignment across stakeholder groups. Interviews further revealed that technology access is a structural exclusion barrier, that older adults are far more heterogeneous than developers assume, and that caregivers bridge this gap by surfacing lived knowledge of older adults' needs that neither developers nor models can capture alone. Together, these findings show that human-centric requirements analysis should engage stakeholder groups explicitly rather than collapsing their perspectives into a single aggregate view. This paper contributes an identification of the key human aspects driving requirement priorities in aged-care digital health and an explainable, human-centric RE framework that combines ML-derived importance rankings with qualitative validation to surface the stakeholder misalignments that inclusive requirements engineering must address.

\end{abstract}




\received{20 February 2007}
\received[revised]{12 March 2009}
\received[accepted]{5 June 2009}

\maketitle
\section{Introduction}
\label{sec:intro}

Digital health software for older adults has become a global priority~\cite{world2021decade}. However, Requirements Engineering (RE) remains insufficiently equipped to address the needs of aged-care systems~\cite{xiao_requirements_2025}. Existing approaches often lack effective human-centric analysis, leading to inadequate requirements prioritisation~\cite{muller_so_2022}. In particular, they overlook the role of human aspects—such as health conditions, living situation, and social context—that shape what stakeholders consider important and why~\cite{grundy_vision_2023}. Understanding which human aspects are most impactful, and for whom, is therefore critical for inclusive and evidence-based RE~\cite{xiao_requirements_2025,noauthor_stakeholders_nodate}. Without such evidence, prioritisation risks being driven by assumptions rather than grounded understanding. Prior work has examined human aspects, but largely relies on expert judgement rather than large-scale user studies with human-in-the-loop validation, limiting its ability to support transparent, stakeholder-aware decisions~\cite{elkady_prioritizing_2024,shamsujjoha_human-centric_2021,hidellaarachchi_effects_2022}.

These limitations highlight the need for approaches that systematically analyse how human aspects influence requirement priorities. Traditional stakeholder engagement relies on manual elicitation and informal interpretation, making it difficult to identify which aspects matter most, how they differ across groups, and where priorities align or conflict~\cite{zainal_exploring_2025,liu_requirements_2016}. This is particularly challenging in aged-care contexts, where variation in health status, technology familiarity, and care conditions is substantial.

Two complementary research streams provide partial solutions but have rarely been integrated. Research on human aspects in RE has shown that factors such as experience, context, and user characteristics influence requirements decisions~\cite{hidellaarachchi_effects_2022,shamsujjoha_better_2025}. In parallel, explainable AI research in software engineering has developed techniques such as SHAP (SHapley Additive exPlanations)~\cite{lundberg_unified_2017} and permutation importance~\cite{tantithamthavorn_explainable_2022,jiarpakdee_empirical_2022,linardatos_explainable_2020} to make predictive models interpretable. However, these techniques have been applied primarily to tasks such as defect prediction and issue classification, rather than to requirements prioritisation or stakeholder analysis.

To address this gap, we conduct an explanatory mixed-methods study that combines survey-based explainable machine learning with semi-structured interviews. Using data from 250 participants across three stakeholder groups—older adults, caregivers, and developers—we identify the human aspects most strongly associated with perceived requirement importance, examine how these patterns differ across groups, and use interviews to explain and validate the results. Our focus is not only on identifying important factors, but on making these factors interpretable and actionable for RE. By linking explainable model outputs with stakeholder perspectives, we aim to support more transparent and stakeholder-aware requirements prioritisation in aged-care digital health.

\noindent\textbf{Contributions.}
This paper makes two contributions. 
(1) \textbf{Empirical stakeholder-aware evidence}: We identify the human aspects that most strongly influence perceived requirement importance in aged-care digital health and provide a comparative analysis across older adults, caregivers, and developers, revealing both shared patterns and substantial misalignment in requirement priorities. 
(2) \textbf{Explainable RE approach}: We propose and evaluate a mixed-methods pipeline that combines explainable machine learning with human-in-the-loop validation, enabling transparent, comparative, and stakeholder-aware requirements prioritisation.

\noindent\textbf{Research Questions}
\begin{itemize}
    \item \textbf{RQ1:} Which human aspects are most important for aged-care digital health software requirements, as identified through explainable ML analysis of survey data?
    \item \textbf{RQ2:} How do these important human aspects differ across older adults, developers, and caregivers?
    \item \textbf{RQ3:} How do interview findings explain, validate, or refine the human-aspect importance patterns identified by explainable ML?
\end{itemize}

\section{Background and Related Work}
\label{sec:related}
% What to report:
% - Human aspects in RE for digital health / aged care.
% - Multi-stakeholder requirements and perception gaps.
% - Explainable ML / SHAP / model-agnostic explanation in software engineering.
% - Mixed-method empirical SE studies that combine quantitative identification + qualitative explanation.
%
% Recommended subsection split:
\subsection{Human Aspects in Aged-Care Digital Health Requirements}
Requirements Engineering (RE) activities such as elicitation, prioritisation, and negotiation are strongly shaped by human aspects rather than being purely technical tasks~\cite{curumsing_emotion-oriented_nodate, hidellaarachchi_effects_2022}. In aged-care digital health, this challenge is amplified: older adults, caregivers, and developers represent heterogeneous stakeholders with different goals, capabilities, and constraints, each likely to emphasise different concerns and interpretations of system needs~\cite{xiao_requirements_2025, noauthor_stakeholders_nodate}. Prior work confirms that human aspects such as personality and experience influence how practitioners engage in RE~\cite{hidellaarachchi_effects_2022, shamsujjoha_better_2025}, but which aspects matter most for which requirement types—and how this varies across stakeholder groups—remains underexplored. 

\subsection{Multi-Stakeholder Requirements Engineering}
A broad body of work demonstrates that stakeholder viewpoints in RE are often complementary but also conflicting, requiring consistency checking, negotiation, and conflict management~\cite{sommerville_viewpoints_1998, nuseibeh_framework_1994, nuseibeh_leveraging_2000, muller_so_2022, elkady_prioritizing_2024}. Subsequent research extended this by developing methods for reasoning about inconsistent viewpoints and resolving conflicts among stakeholder expectations~\cite{schulte_studying_2024, haldar_interpretable_2024}. In digital health for older adults, this challenge is especially acute because end users, developers, and caregivers bring fundamentally different forms of knowledge—shaped by lived experience, technical expertise, and clinical practice, respectively~\cite{shamsujjoha_developer_2024, shamsujjoha_human-centric_2021}. A stakeholder-aware approach is therefore needed not only to capture requirements, but to understand how priorities vary across groups and what those differences imply for the engineering process~\cite{xiao_requirements_2025}.
\subsection{Explainable Analytics for Software Engineering Decisions}
% Position SHAP as a way to explain why aspects matter, not just predict.
There is growing interest in applying explainable methods to interpret predictive models in software engineering~\cite{tantithamthavorn_explainable_2022, jiarpakdee_empirical_2022, wolf_explainability_2019}. Model-agnostic techniques such as SHAP have been shown to reveal the factors driving model outputs, support defect prediction~\cite{lundberg_unified_2017,jiarpakdee_empirical_2022}, explain issue classification~\cite{schulte_studying_2024}, and estimate maintenance effort~\cite{haldar_interpretable_2024}. Both post-hoc explainers and inherently interpretable models~\cite{girard_inclusive_nodate} aim to improve transparency in software analytics, though they are not equivalent~\cite{wang_model_2026, turbe_evaluation_2023}. Crucially, these techniques shift analytics from opaque ranking toward interpretable, stakeholder-centric reasoning~\cite{linardatos_explainable_2020, girard_inclusive_nodate}—making them well-suited for requirements analysis, where understanding \emph{why} certain factors matter is as important as knowing \emph{which} factors rank highest.

\section{Study Design}
\label{sec:design}
% What to report:
% - Overall mixed-method design: explanatory sequential is best.
% - Phase 1: survey + ML + SHAP.
% - Phase 2: interviews for explanation/validation.
% - Rationale for mixing methods.
%
% Suggested wording:
% "We use an explanatory sequential mixed-method design, where quantitative findings
% from a cross-sectional survey are first analysed using ML+SHAP, followed by interviews
% to explain and validate the resulting rankings."
\begin{figure}[h]
  \centering
  \includegraphics[trim={0cm 5cm 0cm 4cm}, clip, width=1.05\linewidth]{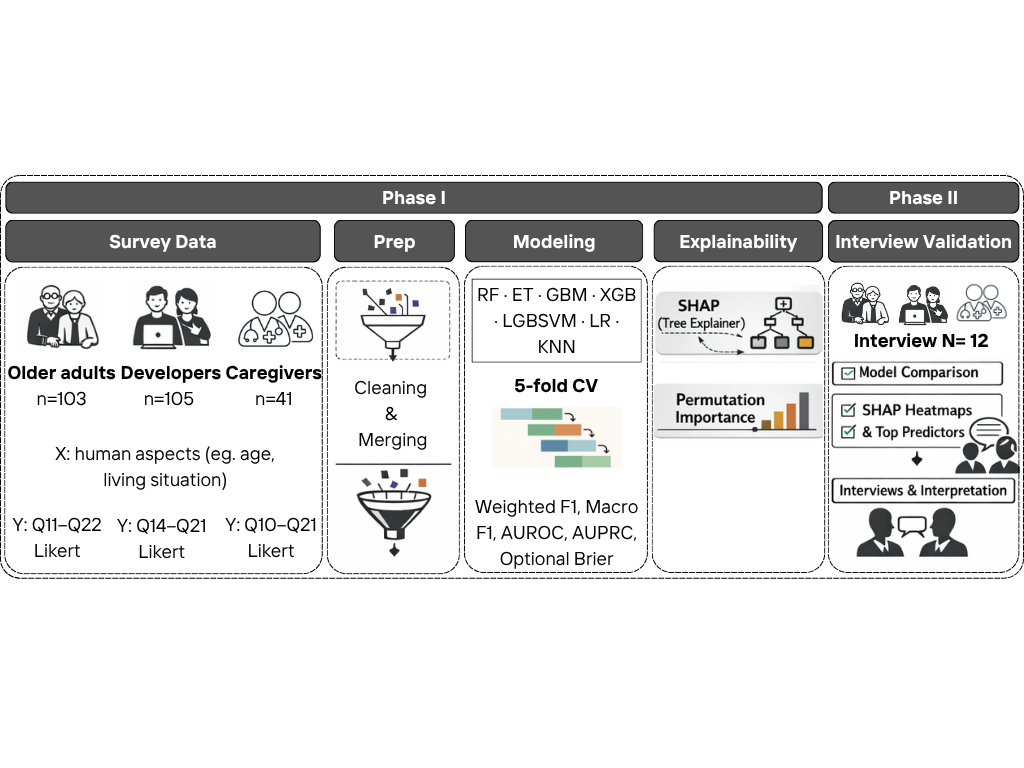}
  \caption{Method Flow for study design.}
  \vspace{2pt}
    {\scriptsize \textbf{Abbreviations:} CV: Cross-Validation. RF: Random Forest. ET: Extra Trees (Extremely Randomised Trees). GBM: Gradient Boosting Machine. XGB: XGBoost (Extreme Gradient Boosting). LGBSVM: LightGBM-SVM (A hybrid model using LightGBM for feature selection or Gradient Boosting Trees with a Support Vector Machine layer). LR: Logistic Regression. KNN: k-Nearest Neighbours.}
  \label{fig:method}
\end{figure}
As summarised in Figure~\ref{fig:method}, we adopted an explanatory sequential mixed-methods design to investigate how human aspects influence requirements prioritisation in aged-care digital health software. Phase 1 applied supervised machine learning with model-agnostic explanation methods—SHAP~\cite{lundberg_unified_2017} and permutation importance~\cite{noauthor_permutation_nodate}—to analyse survey data from older adults, caregivers, and developers, identifying the human aspects most strongly associated with requirement-priority outcomes. And then, in Phase~2, we used semi-structured interviews to validate and extend Phase~1 findings and explain the underlying patterns.

This mixed-methods approach combines data-driven analysis
with qualitative interpretation, as guidelines suggested~\cite{sandelowski_mapping_2012}. The quantitative phase systematically identified influential predictors across outcomes and stakeholder groups, while SHAP and permutation importance ensured interpretability by revealing which human aspects drove model behaviour. The qualitative phase then provided explanatory validation, assessing whether the identified patterns aligned with stakeholder experiences and clarifying why differences emerged. Together, these phases enable a more interpretable and stakeholder-aware understanding of requirements prioritisation.

\paragraph{\textbf{Research Context and Participants}}

Three stakeholder groups were recruited to capture distinct yet interdependent perspectives on aged-care digital health: older adults as primary end-users, caregivers (formal and informal) as the care support layer, and software developers as system architects. Participants were sourced through relevant communities (e.g., senior online chatting groups), professional networks, and organisations. Eligibility required participants to be 18 or older and to self-identify with one of the target groups; older adults additionally met a later-life user definition, caregivers had formal or informal care experience, and developers had relevant digital systems experience. All participants provided informed consent following receipt of a study information statement; ethics approval was obtained from the relevant institutional committee. Background data were collected across groups on age, health context, occupation, living situation, device use, and other human aspects relevant to Digital Health (DH) engagement.

\paragraph{\textbf{Mixed-Methods Design Rationale}}

We adopted an explanatory sequential mixed-methods design to investigate the intersection of human aspects and DH requirement priorities~\cite{sandelowski_mapping_2012, chen_digital_2023}. This two-phase approach (see Figure~1) was designed to combine the statistical power of machine learning with the interpretive depth required to understand stakeholder motivations. Given the inherent complexity of aged-care contexts, traditional linear models may fail to capture non-linear relationships and high-order interactions between human aspects. In Phase~1, supervised machine learning (such as XGB) systematically ranked which human aspects most strongly influenced requirement outcomes, with SHAP and Permutation Importance providing transparent, multi-faceted explanations that standard regression cannot offer. However, quantitative importance scores alone cannot explain the human rationale behind them. Following established guidelines~\cite{morse_significance_1995, Barroga2022}, 12 participants were selected across all three stakeholder groups to ensure cross-group comparison remained feasible. Phase~2 therefore conducted follow-up semi-structured interviews (N=12) to validate and contextualise the Phase~1 findings across three dimensions: \textit{convergence} (whether top SHAP predictors aligned with stakeholders' lived experiences), \textit{divergence} (why certain aspects mattered for one group but not others), and \textit{interpretability} (unpacking results where high predictive power lacked an obvious theoretical explanation). This sample size is consistent with established guidance for explanatory sequential designs, where the qualitative phase is scoped to explain and validate quantitative findings rather than achieve independent saturation~\cite{creswell_research_2018}. Theoretical sufficiency was reached when recurring themes—technology access barriers, developer--user perspective gaps, and within-group heterogeneity—ceased to introduce new explanatory content across successive interviews. Together, the two phases ensure findings are both statistically robust and practically grounded for DH requirements engineering.
% What to report:
% - Why survey first?
% - Why interviews second?
% - What the interviews validate: convergence, divergence, unexpected SHAP results.

\subsection{Phase 1: Survey Data Curation, Model Fitting, and Explainability Analysis}
\label{sec:phase1}

Phase~1 comprised three sequential procedures: (\textit{i}) survey data curation, in which responses from the three stakeholder groups were cleaned, harmonised, and transformed into analysis-ready datasets; (\textit{ii}) model fitting, in which supervised learning models were trained to predict requirement-priority outcomes from human-aspect predictors; and (\textit{iii}) explainability analysis, in which SHAP and permutation importance were used to identify the predictors that contributed most strongly to model behaviour. This procedural flow follows an ASE-style empirical design in which data preparation, model construction, and model interpretation are treated as distinct but connected stages of analysis. 

% We conceptualised \emph{human aspects} as the person- and context-specific characteristics that shape how stakeholders evaluate and prioritise DH requirements. These served as our independent variables ($X$) and encompassed demographics (e.g., age, living situation), health conditions (eg. has chronic disease, has memory declines, etc.), technological proficiency, and device ownership.

% The outcome variables ($Y$) consisted of ordinal Likert-scale responses representing perceived requirement importance and design preferences. As shown in Figure~\ref{fig:method}, these outcomes were mapped to specific survey instruments: items Q12--Q22 for older adults, Q14--Q21 for developers, and Q10--Q21 for caregivers. 
We conceptualised \emph{human aspects} as the person- and context-specific characteristics that shape how stakeholders evaluate and prioritise DH requirements. These served as our independent variables $\mathbf{x} = \{x_1, x_2, \dots, x_n\}$ and encompassed demographics (e.g., age, living situation), health conditions (e.g., chronic disease, memory decline), technological proficiency, and device usage.

The outcome variables $\mathbf{y} = \{y_1, y_2, \dots, y_m\}$ consisted of ordinal Likert-scale responses representing perceived requirement importance and design preferences. As shown in Figure~\ref{fig:method}, these outcomes were mapped to specific instruments from surveys with participants. Formally, for each outcome $y_j \in \mathbf{y}$, we train a model $f_j : \mathbb{R}^n \rightarrow \mathcal{Y}_j$ such that $\hat{y}_j = f_j(x_1, \dots, x_n)$, where $\mathcal{Y}_j$ is the discrete Likert scale of outcome $j$. To identify which human aspects drive each prediction, we apply SHAP~\cite{lundberg_unified_2017}, which decomposes $\hat{y}_j = \phi_0^{(j)} + \sum_{i=1}^{n} \phi_i^{(j)}(x_i)$, where $\phi_0^{(j)}$ is the baseline prediction and $\phi_i^{(j)}$ quantifies the marginal contribution of $x_i$ to outcome $y_j$. Global importance is summarised as the mean absolute SHAP value across all $N$ participants, $\bar{\phi}_i^{(j)} = \frac{1}{N} \sum_{k=1}^{N} |\phi_i^{(j)}(x_i^{(k)})|$, enabling systematic cross-outcome comparison of human aspect importance across all $m$ requirement priority themes.

While survey instruments were tailored to each group's expertise, semantically equivalent variables were harmonised to enable cross-group comparison. This variable set were published in a systematic review of DH adoption~\cite{xiao_requirements_2025} and stakeholder-centred design literature to ensure coverage of the most plausible drivers of requirement prioritisation. Critically, all predictors $x_i \in \mathbf{x}$ were benchmarked against the older adult profile: developers and caregivers reported the human-aspect attributes of their target users or patients rather than their own, enabling all stakeholder data to be pooled into a single unified feature space for model fitting. Table~\ref{tab:xy_themes} delineates these predictors across requirement themes, highlighting categories with significant cross-outcome predictive power.

\begin{table}[t]
\centering
\caption{Taxonomy of Human Aspects for Aged-Care RE: predictor categories (X: Human aspects) and outcome themes (y: Perceived requirements importance).}
\label{tab:xy_themes}
\resizebox{\linewidth}{!}{%
\begin{tabular}{p{0.3cm} ll}
\hline
\textbf{Axis} & \textbf{Theme} & \textbf{Included variable(s)} \\
\hline

X & Demographic & Age, gender \\
  & Health conditions & Self-reported health conditions \\
  & Past job & Engineering, management, health, etc.  \\
  & Living situation & Living alone or with others \\
  & Device used & Smartphone, tablet, laptop, desktop, etc. \\
  & Social engagement & Social contact software types, frequency of interaction \\
  & Technology preferences & App usage tendencies \\
  & Help Received  & Install/tech helper \\

\hline

y & Current AI-based DH functions & Health monitoring, sleep tracking, etc. \\
  & Current UX & Easy to use, quickly responsive, etc. \\
  & DH Data awareness & Data consent awareness and experience \\
  & Expected AI-based DH functions &Health monitoring, sleep tracking, etc. \\ 
  & Expected UX & Easy to use, quickly responsive, etc.\\
  & Data collection attitude & Comfort level of sharing data \\
  & Personalisation & Personalised UI, reminder, etc.\\

\hline
\end{tabular}%
}
\end{table}

\subsubsection{Data Curation and Pre-processing}\label{sec:data-curation}

Our data curation pipeline follows established empirical standards in software engineering, transforming raw survey instruments into high-fidelity observation matrices~\cite{kitchenham_preliminary_2002, kitchenham_evidence-based_2004}. We deployed a large-scale online survey via Qualtrics, capturing a diverse initial sample of $N=265$ across three stakeholder groups: developers ($n=113$), older adults ($n=111$), and caregivers ($n=41$).To ensure internal validity and mitigate the impact of low-effort responses, we applied a rigorous four-stage screening protocol: 1) \textbf{Completion Check:} Removal of incomplete survey sessions; 2)  \textbf{Attention Check:} Exclusion of 14 participants who failed embedded trap questions; 3) \textbf{Deduplication:} Verification of unique respondents via IP addresses, demographic cross-referencing, and qualitative consistency; 4) \textbf{Response Quality:} Removal of two "speeders" and "straight-liners" whose response patterns indicated a lack of engagement.The final analytic sample comprised $249$ valid responses (105 developers, 103 older adults, and 41 caregivers). 

We then encoded categorical predictors—including device-use context, living situation, and health-related factors—into structured feature matrices suitable for non-parametric modelling. To address the trade-off between feature depth and comparative breadth, we curated two distinct dataset configurations:

\textbf{Harmonised Cross-Group Dataset.} To enable cross-group analysis, we aligned semantically equivalent variables across the three instruments, yielding a shared feature space with 19 predictors and 45 outcomes. We also added a categorical \textit{Stakeholder Type} variable to capture group identity in the combined models.

\textbf{Stakeholder-Specific Datasets.} To retain group-specific human aspects (e.g., seniors' private human aspects), we maintained separate datasets for each stakeholder. This configuration supports high-fidelity within-group modelling and post-hoc interpretation. We treat our stakeholder-specific models as explanatory instruments rather than purely predictive ones.  For the caregiver-specific models, SMOTE~\cite{chawla_smote_2002} oversampling was applied within each training fold to mitigate the effects of the smaller sample size on class distribution. 
\subsubsection{Model Fitting}
To identify the predictors most strongly associated with each requirement-priority outcome, we compared several supervised machine learning models, including Random Forest~\cite{breiman_random_2001}, Extra Trees~\cite{geurts2006extremely}, Gradient Boosting~\cite{friedman_greedy_2001}, XGBoost~\cite{chen2016xgboost}, Lasso regression models~\cite{hosmer2013applied}, and k-nearest neighbours (KNN)~\cite{fix1985discriminatory}. These models were selected to compare tree-based ensembles, boosting methods, linear models, kernel methods, and instance-based learning within a common evaluation framework.

\paragraph{\textbf{Model Validation Strategy}}
We fitted supervised learning models to predict each requirement-priority outcome from human-aspect predictors. As the goal was to identify robust and interpretable patterns rather than optimise a single benchmark score, we used stratified 5-fold cross-validation (CV)~\cite{stone_cross-validatory_1974} instead of a fixed train--test split. In each fold, approximately 80\% of data were used for training and 20\% for testing, while preserving the distribution of ordinal outcome categories.

To reduce over-parameterisation, we applied feature importance filtering, retaining only the most influential predictors for the final analysis. To address class imbalance in the caregiver dataset (n=41), we applied SMOTE~\cite{chawla_smote_2002} to the training data within each fold, after splitting, to avoid data leakage.

\paragraph{\textbf{ML Performance Metrics}}
We report weighted F1~\cite{van_rijsbergen_theoretical_1977} to account for class imbalance, macro F1~\cite{hirschman1998evolution} to ensure minority-class performance, and AUROC~\cite{mcdermott_closer_2025} and AUPRC~\cite{hutchison_area_2013} to assess discrimination and precision--recall trade-offs. These metrics ensure that models are sufficiently reliable for interpretation rather than optimised purely for prediction.  

\subsubsection{Explainability Analysis}
\label{sec:explainability}

After model comparison, we used the best-performing model for each outcome for explanation analysis. In practice, Random Forest showed strong and stable performance across outcomes and supported efficient tree-based SHAP computation, so it was used as the primary model for interpretation. We computed SHAP values with TreeExplainer and used mean absolute SHAP values to obtain global predictor rankings for each outcome.

These rankings produced outcome- and stakeholder-specific importance profiles, which we visualised with heatmaps and beeswarm plots. Heatmaps enabled compact comparison of top predictors across outcomes and groups, while beeswarm plots showed the spread and direction of effects for selected predictors. As a robustness check, we also computed permutation importance by randomly permuting each predictor over 20 repeats and measuring the average decrease in weighted F1. We used SHAP and permutation importance as complementary explanation methods to understand why the selected models behaved as they did.

The outputs of this phase were: (\textit{i}) global top-predictor rankings for each requirement-priority outcome, (\textit{ii}) comparative importance profiles across older adults, developers, and caregivers, and (\textit{iii}) a set of convergent, divergent, and unexpected predictor patterns used to guide the interview protocol in Phase~2. The anonymised code and data are available in the online supplementary materials.
\subsection{Phase 2: Interview-Based Validation}
\label{sec:phase2}

Phase~2 provides the interpretive depth necessary to unpack the "black-box" findings from the Phase~1 predictive models. While the quantitative analysis identifies which human aspects correlate with requirement priorities, it cannot inherently explain the underlying socio-technical rationale. We conducted semi-structured interviews ($N=12$) to determine if the model-derived importance rankings are contextually meaningful and to refine our understanding of unexpected SHAP results. Consequently, Phase~2 functions as a targeted explanatory follow-up, shifting the focus from statistical association to participant-driven causality.

We employed a purposive sampling strategy (selects participants based on specific characteristics relevant to the research question, ensuring an information-rich sample~\cite{Palinkas2013}) to recruit participants from Phase~1 cohort—older adults, caregivers, and software practitioners with expertise in SE, HCI, Digital Health, and RE—to ensure direct relevance to the quantitative patterns identified in Phase~1. The full demographics of interviewees are available in the online materials.

Participants were recruited through professional networks and referrals to balance continuity with the Phase~1 cohort and variation in professional experience. Selection prioritised the ability to provide informed reflection on the quantitative findings from lived, professional, or design-oriented perspectives. The interviews were intended for analytic validation and contextual explanation rather than statistical generalisation, assessing whether the identified human aspects are realistic, interpretable, and actionable for aged-care RE practice.

\paragraph{\textbf{Interview Sampling and Protocol}}

The interview protocol was derived directly from the Phase~1 outputs. Specifically, the interview guide was informed by the top-ranked human aspects identified through SHAP and permutation importance analysis, as well as outcomes where importance patterns diverged across stakeholder groups. This design ensured that the qualitative phase remained tightly coupled with the quantitative results.

Each semi-structured interview addressed four main topics~\cite{groen_guidelines_2021}. Participants were asked to reflect on which human aspects they considered most relevant to aged-care digital health, how such aspects relate to requirement priorities in practice, which aspects they felt were commonly overlooked in development processes, and what they believed should inform future elicitation and design decisions. The full interview guide is available in the online supplementary materials.

Followed by established empirical software engineering guidelines, after core quantitative patterns were identified, subsequent qualitative analysis focuses on explaining those patterns, assessing their practical significance, and clarifying the conditions under which they hold~\cite{sandelowski_mapping_2012,kitchenham_preliminary_2002}.

\paragraph{\textbf{Qualitative Analysis}}
We analysed the interview data using a hybrid thematic analysis approach. The coding framework was deductively anchored in the Phase~1 outputs—specifically the SHAP-derived feature rankings and cross-group importance variances—while remaining open to inductive theme development for aspects not captured by the quantitative models.

Followed by meta analysis guidelines~\cite{forero_ten_2019, hansen_how_2022, aguinis_meta-analytic_2011}, the analysis proceeded in three stages: 1) \textbf{Initial Coding}: We established a framework based on \textit{sensitizing concepts} from Phase~1, focusing on the confirmation of high-ranking predictors and explanations for stakeholder divergences; 2)\textbf{Inductive Expansion:} We generated new codes for unanticipated issues raised by participants, such as hidden contextual tensions between caregivers and developers or difficulties in operationalizing specific human aspects; 3)\textbf{Thematic Consolidation:} Related codes were aggregated into higher-level themes. To ensure inter-coder reliability, the first author coded all 12 transcripts; three co-authors independently reviewed the resulting themes and raised challenges, which were resolved through iterative discussion until consensus was reached.

The authors conducted internal calibration sessions to refine code definitions and resolve interpretive disagreements. This iterative reconciliation process was employed to ensure inter-coder consistency while preserving the underlying thematic nuance. By formalising these review procedures, we ensure that the qualitative analysis serves as a rigorous, systematic validation of the Phase~1 results rather than an informal interpretative layer.

\paragraph{\textbf{Integration of Quantitative and Qualitative Findings}}
We integrated the interview findings with the explainability results through explanatory linking, mapping Phase~2 themes back to the Phase~1 importance patterns. This allowed us to assess whether the quantitative findings were \emph{confirmed}, \emph{explained}, \emph{challenged}, or \emph{refined}. In this way, Phase~2 used stakeholder interpretation to validate and contextualise them, showing where the models were well grounded, misleading, or incomplete.

\section{Results}
\label{sec:results}
% Best organization: by RQ, not by method.
% Each RQ subsection should integrate quant + qual where relevant.
Results are presented across three parts: RQ1 identifies the dominant human aspects, RQ2 reveals how these diverge across stakeholder groups, and RQ3 uses interview evidence to validate and explain findings in RQ1 and RQ2. Prior to importance analysis, we verified model reliability: AUROC ranges from 0.614 to 0.907 and AUPRC from 0.363 to 0.541 across seven requirement-priority groups (Appendix B), with weighted F1 between 0.484 and 0.807; six groups exceed AUROC 0.78, confirming adequate discriminatory power. (Details in online materials)
\subsection{RQ1: Important Human Aspects Identified by ML+SHAP}
% What to report:
% - Overall top human aspects.
% - Their SHAP magnitude / contribution strength.
% - Whether effects are positive/negative if meaningful.
% - One figure: overall SHAP summary plot.
% - One table: ranked aspects with descriptions.

\begin{figure}[ht]
  \centering
  \includegraphics[ width=\linewidth]{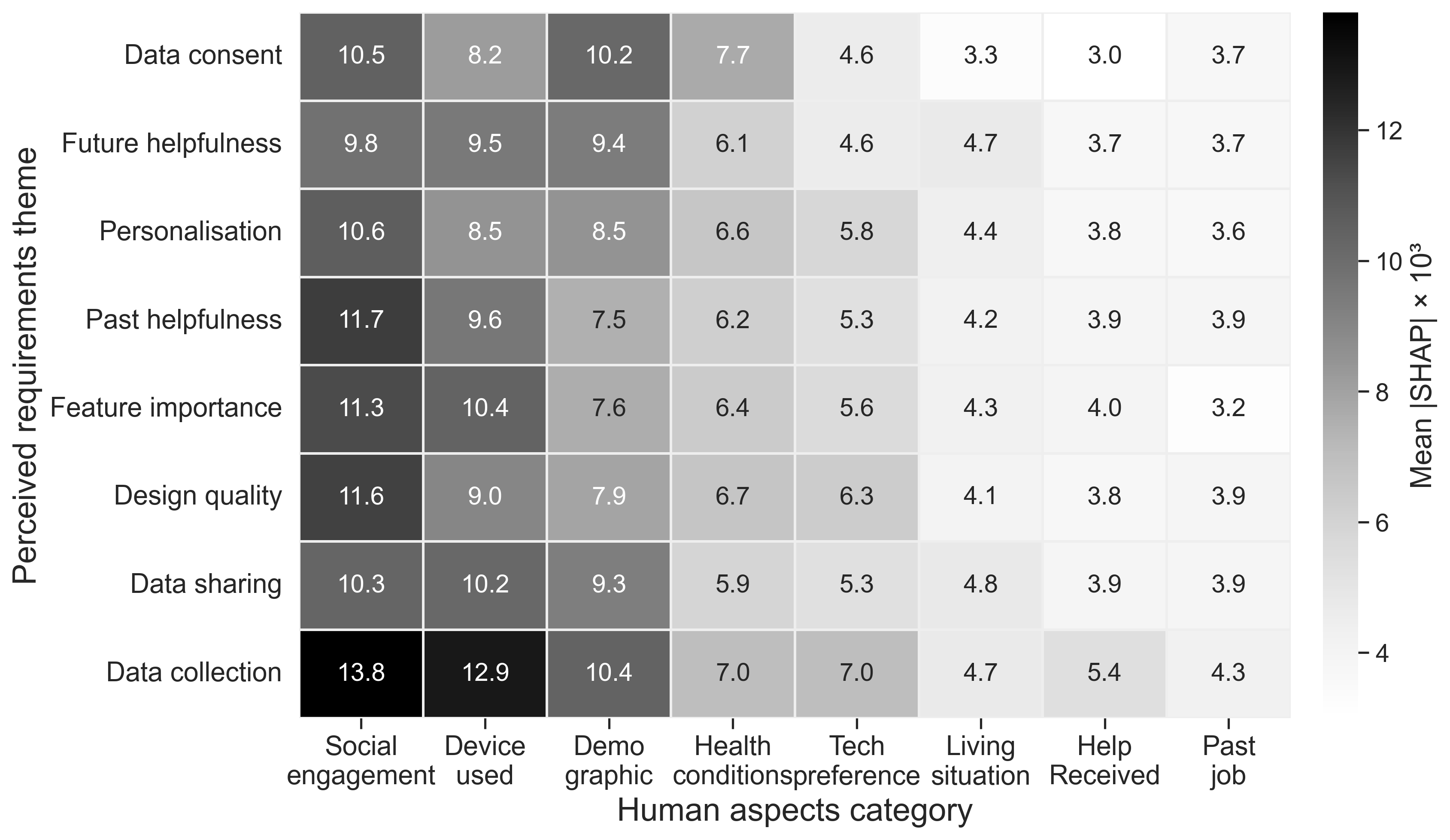}
  \caption{Human aspects VS perceptions heatmap}
  \label{fig:rq1_heatmap}
\end{figure}
\begin{figure}[h]
  \centering
  \includegraphics[trim={0cm 0cm 3cm 0cm}, clip, width=0.95\linewidth]{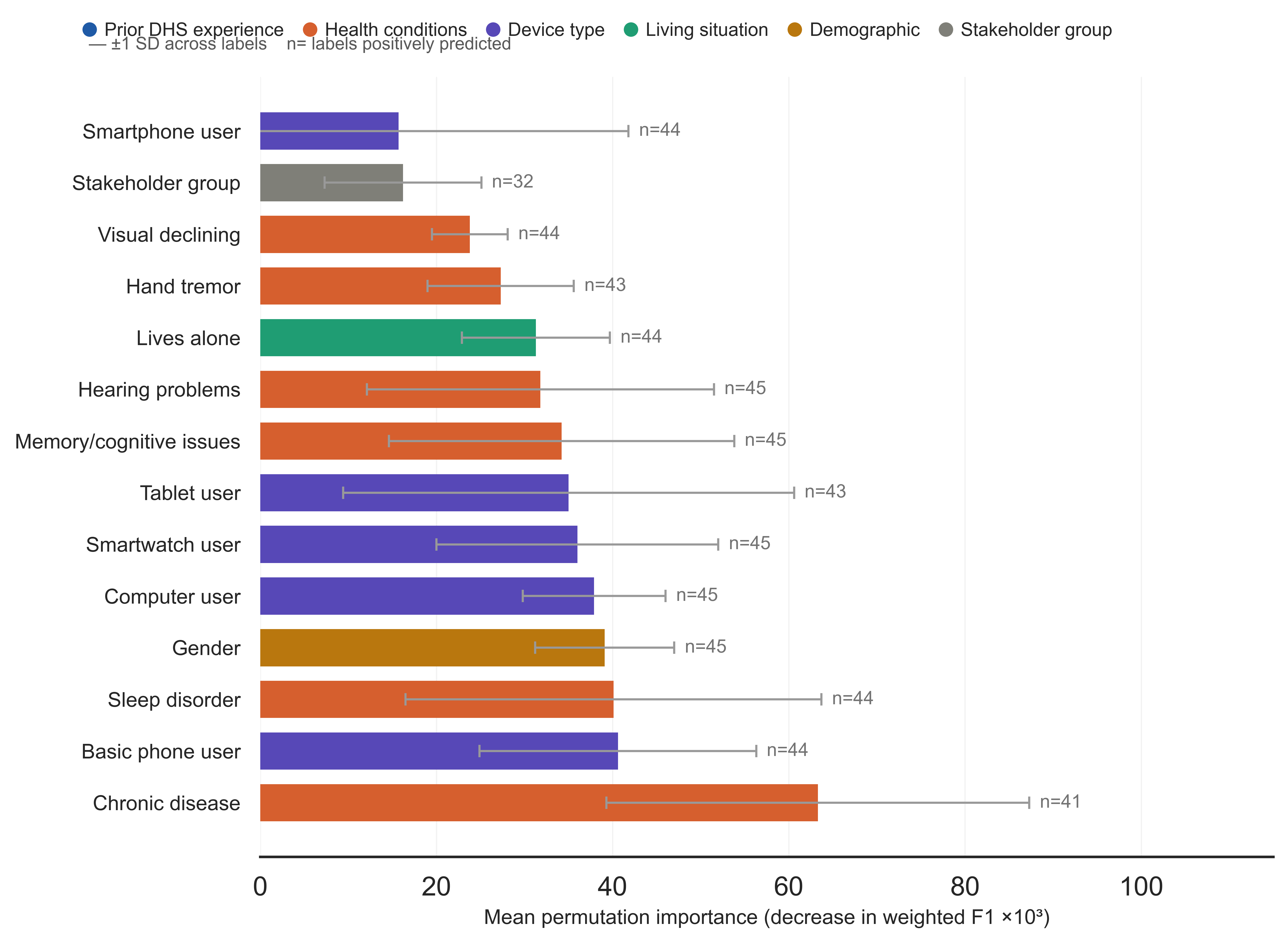}
  \caption{Top Confirmed Human aspect Predictor}
  \label{fig:rq1_top_predictors}
  \vspace{2pt}
    {\scriptsize Mean permutation importance across all labels where the predictor is a confirmed positive signal. Error bars = ±1 SD. Only predictors confirmed by both SHAP and permutation importance (perm > 0) are shown.}
\end{figure}

To address RQ1, we analysed the global importance of human-aspect predictors across all requirement-importance outcomes using SHAP, and cross-checked the resulting rankings using permutation importance. Figure~\ref{fig:rq1_heatmap} presents the mean absolute SHAP importance of human factor themes ((X)) across all requirement-priority outcomes ((Y)), and Figure~\ref{fig:rq1_top_predictors} identifies the highest-ranking individual predictors within the combined dataset.

Social engagement, device used, and demographic characteristics are the dominant predictors of digital health requirement priorities.
\textbf{Social engagement} yields the highest SHAP contributions for nearly every outcome, most strongly for \emph{data collection} (13.8), \emph{perceived usefulness} (11.7), and \emph{current} and \emph{desired UX requirements} (11.6 and 11.3, respectively)—indicating that social context shapes both data governance and experiential priorities alike. \textbf{Device used} consistently ranks second, with pronounced effects on \emph{data collection} (12.9), \emph{desired UX requirements} (10.4), and \emph{data sharing} (10.2). \textbf{Demographic factors} complete the top tier, with strong contributions to \emph{DH data consent} (10.2), \emph{data collection} (10.4), and \emph{future AI-based DH functions} (9.4).

Health conditions and technology preferences exert consistent secondary influence, with particular salience for data-sensitive outcomes. Both categories peak at \emph{data collection} (7.0 each), with \textbf{health conditions} further elevating \emph{current UX requirements} (6.7) and \emph{personalisation} (6.6), suggesting that self-reported health status jointly shapes users' data sensitivity and adaptation expectations. \textbf{Technology preferences} show a parallel pattern, particularly for \emph{current UX requirements} (6.3) and \emph{desired UX requirements} (5.6). \textbf{Living situation}, \textbf{support received}, and \textbf{past job} contribute at lower magnitudes, though their recurrence across outcomes warrants their retention as contextual covariates.
\begin{findingbox}
\stepcounter{findingcount}
\noindent \textbf{Finding \arabic{findingcount}:} Human-centric requirements in aged-care digital health are not driven by demographic factors alone; rather, they emerge from a complex interplay of \textbf{health conditions, tech proficiency, and social context}. By modelling these as human aspects, developers can move beyond "one-size-fits-all" inclusive design toward evidence-based inclusive RE.
\end{findingbox}

At the level of individual predictors, Figure~\ref{fig:rq1_top_predictors} illustrates that chronic disease and sleep disorder emerged as the most consistently influential variables. These were followed by a secondary tier comprising basic phone use and gender, with various device-use variables rounding out the lower-ranked predictors. Other highly ranked predictors included \textbf{sleep disorder}, \textbf{gender}, \textbf{memory/cognitive issues}, \textbf{hearing problems}, \textbf{living alone}, \textbf{hand tremor}, \textbf{visual problems}, and \textbf{stakeholder group}. Taken together, these results show that the influential human aspects are not limited to demographics or usability preferences alone, but span health, technology-use context, social context, and prior experience.

These findings also reveal meaningful variation across outcome themes. \emph{Data collection} was the most strongly structured outcome overall, receiving the highest SHAP values across nearly all predictor categories. In contrast, outcomes such as \emph{consent awareness}, \emph{expected usefulness}, and \emph{personalisation} were still influenced by multiple human-aspect categories, but with lower overall magnitudes. This suggests that attitudes toward data-related functionality may be especially sensitive to differences in participants' health, device-use, and social-context profiles. 

Furthermore, when the stakeholder group was separated into separate indicators, the Developer group showed the highest overall predictive importance, followed by caregivers and older adults, based on mean permutation importance across all outcomes.

\begin{findingbox}
\stepcounter{findingcount}
\noindent \textbf{Finding \arabic{findingcount}:} ``Basic phone user" likely serves as a proxy for \textbf{lower technical proficiency} or \textbf{socioeconomic constraints}, suggesting that hardware familiarity is a primary determinant of how stakeholders perceive digital health needs. 
\end{findingbox}

Because the SHAP results reported here are based on mean absolute contributions, they indicate \emph{strength of influence} rather than whether a predictor increased or decreased the perceived importance of a requirement. We therefore interpret the results as identifying the most influential human aspects overall, rather than as establishing directional effects. Directionality for selected outcomes is examined later using beeswarm plots and interview-based interpretation.

Since we report mean absolute SHAP values, our analysis identifies the most influential human aspects (\emph{strength of influence}) rather than directional impacts. We address directionality and causality for specific outcomes later through beeswarm plots and interview data.
\begin{summarybox}
\textbf{Answer to RQ1:} The most important human aspects can be grouped into four categories: \textbf{technology proficiency-related}, \textbf{health condition-related}, \textbf{care context-related}, and \textbf{demographic} factors. In particular, the explainable ML analysis identified \emph{device used} and \emph{technology preferences}, \emph{chronic disease} and \emph{sleep disorder}, \emph{social engagement}, and \emph{gender} and \emph{age range} as the most consistently influential individual predictors. These results indicate that requirement-importance judgments in aged-care digital health software are shaped by a combination of health status, prior experience, technology context, and social context, rather than by demographic factors alone.
\end{summarybox}

\subsection{RQ2: Differences Across OA, Developers, and Caregivers}

\begin{table}[ht]
\caption{Top-5 important human aspects for OA digital health requirements, as reported by three stakeholder groups}
\label{tab:top5_group}
\centering
\scriptsize
\begin{tabular}{llll}
\hline
\textbf{Rk} & \textbf{Older Adults} & \textbf{Dev} & \textbf{Caregiver} \\
\hline
\#1 & \textbf{Has chronic disease --} & \textbf{Target user (OA) Age --} & Caregiver role (informal vs formal) *\\
\#2 & Computer user + & \textbf{Has chronic disease --} & \textbf{Lives w/ family *} \\
\#3 & Online social contact + & Data awareness + & Sleep disorder --\\
\#4 & \textbf{living w/ family *} & Target user (OA) Gender *& Hearing problem --\\
\#5 & \textbf{Age --} & Emotion * & \textbf{Patient (OA) Age --} \\
\hline
\end{tabular}
\vspace{4pt} % Slightly more space for readability
\begin{flushleft}
\footnotesize 
\textbf{Bold terms} indicate human aspects shared across two or more stakeholder groups as identified by the SHAP analysis. Directional impact symbols: ``+" positive overall effect on requirement priority; ``--" negative overall effect; ``*" effect is non-monotonic or context-dependent and is discussed further in Section~\ref{sec:rQ3}.
\end{flushleft}
\end{table}

\begin{figure}[h]
  \centering
  \includegraphics[ width=\linewidth]{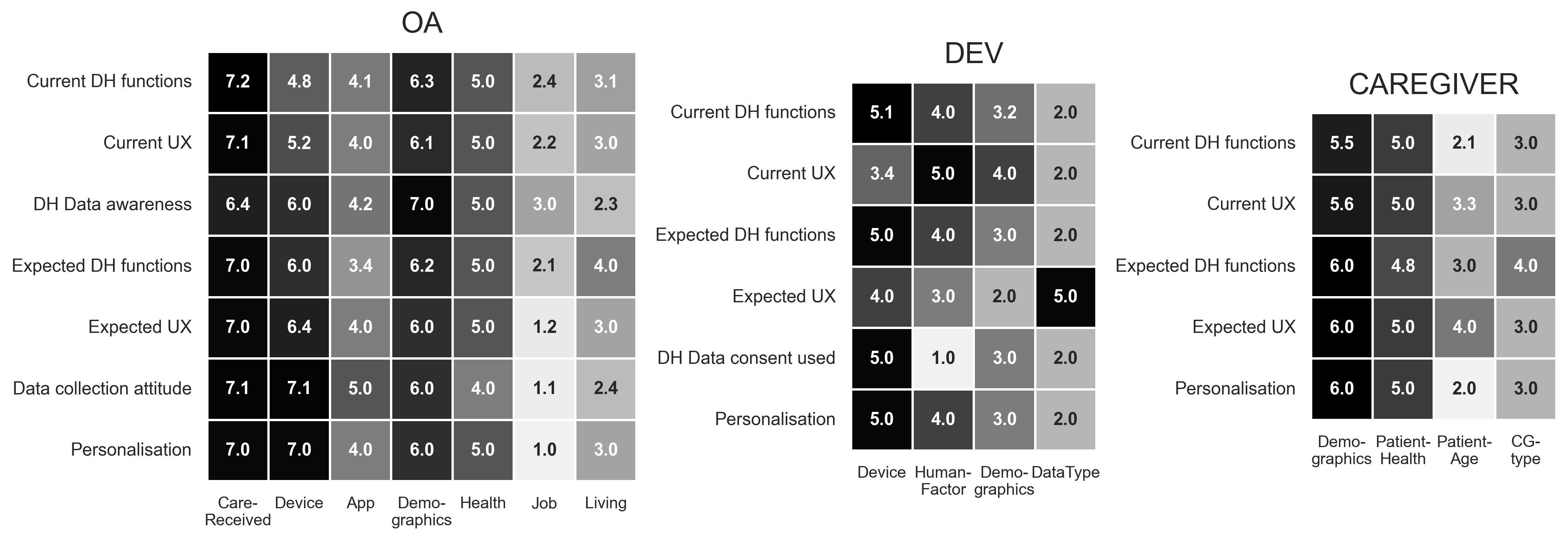}
  \caption{Predictor Category Rankings by Outcome Theme and Stakeholder Group}
  \label{fig:rq2_rank}
  \vspace{2pt}
    {\scriptsize Rank = importance within row (highest = strongest). Dark = high rank. DH: Digital health. Social: Social active levels. CG: Caregivers}
\end{figure}

Figure~\ref{fig:rq2_rank} compares predictor-category rankings across stakeholder groups, and Table~\ref{tab:top5_group} reports the top confirmed individual predictors for each group.
\begin{findingbox}
\stepcounter{findingcount}
\noindent \textbf{Finding \arabic{findingcount}:} Stakeholder groups share only a small subset of influential predictors, with group-specific factors dominating each profile.
\end{findingbox}
As shown in Table~\ref{tab:top5_group}, \textit{chronic disease} and \textit{living arrangement} are the only predictors shared across two or more groups, while the remaining top-ranked predictors are largely group-specific. Older adults are distinguished by ``computer use", ``online social contact", and ``age"; developers by ``target user age", ``data-collection frequency", and ``gender"; and caregivers by ``caregiver role type", ``patient sleep disorder", and ``hearing impairment". This divergence indicates that stakeholders' role shapes which human aspects are deemed important for requirements priority judgement.

\begin{findingbox}
\stepcounter{findingcount}
\noindent \textbf{Finding \arabic{findingcount}:} Predictor-category rankings reveal a perspective gap between stakeholders. Overall, caregiver importance fell \textbf{between developers and seniors} on average, suggesting that caregivers may help bridge the gap between the two groups.
\end{findingbox}

To further explore the results in figure~\ref{fig:rq2_rank} that \textit{stakeholder type} is a key predictor in the combined dataset, figure~\ref{fig:rq2_rank} further shows that the relative importance of predictor categories differed systematically across the three groups. For OA, the highest-ranked categories were \textbf{contact device}, \textbf{device type}, and \textbf{demographic factors}, followed by health and living-situation variables. For developers, the dominant categories were \textbf{feature-related variables}, \textbf{platform or demo context}, and \textbf{data-type or software-characteristic factors}. For caregivers, the most important categories were \textbf{prior DH experience}, \textbf{health conditions}, and \textbf{living or patient-context variables}. This comparison reveals a clear misalignment: OA and caregivers were influenced more strongly by lived context, health, and use conditions, whereas developers prioritised more design- and feature-oriented aspects.

% The per-group model performance results in Figure~\ref{fig:rq2_f1} provide an additional perspective on these differences. Caregiver models achieved the highest weighted F1 scores across nearly all outcome themes, including \emph{past helpfulness}, \emph{design quality}, and \emph{data consent}. Developer models performed best for \emph{data collection}, while OA models showed more moderate performance across most outcome themes. These differences suggest that some stakeholder groups exhibited more internally consistent requirement-priority patterns than others, which in turn affected the ease with which those patterns could be modelled. In particular, the stronger caregiver performance may indicate that caregiver judgments were more structured around a smaller set of influential human aspects.
\begin{summarybox}
\textbf{Answer to RQ2:}
Human aspects that shape digital health requirements priorities \textbf{differ substantially} across older adults, developers, and caregivers; only \textbf{chronic disease and living situation are consistently influential across groups}. Older adults prioritise technology access and lived experience, caregivers emphasise patient context and health, and developers focus on design and functionality. For requirements engineering, this matters: a process that draws on a single stakeholder group risks systematically overlooking what the others consider important.
\end{summarybox}

\subsection{RQ3: Interview Validation and Explanatory Themes}\label{sec:rQ3}
% What to report:
% - Which SHAP findings were confirmed.
% - Which were nuanced/refined.
% - Which were challenged.
% - 3--5 themes only, tightly linked to quantitative findings.
%
% Example reporting pattern:
% "Theme 1 explains why caregivers prioritise X more than developers..."
% "Theme 2 clarifies why aspect Y is important but under-recognised by models..."

\begin{figure}[h]
  \centering
  \includegraphics[ width=\linewidth]{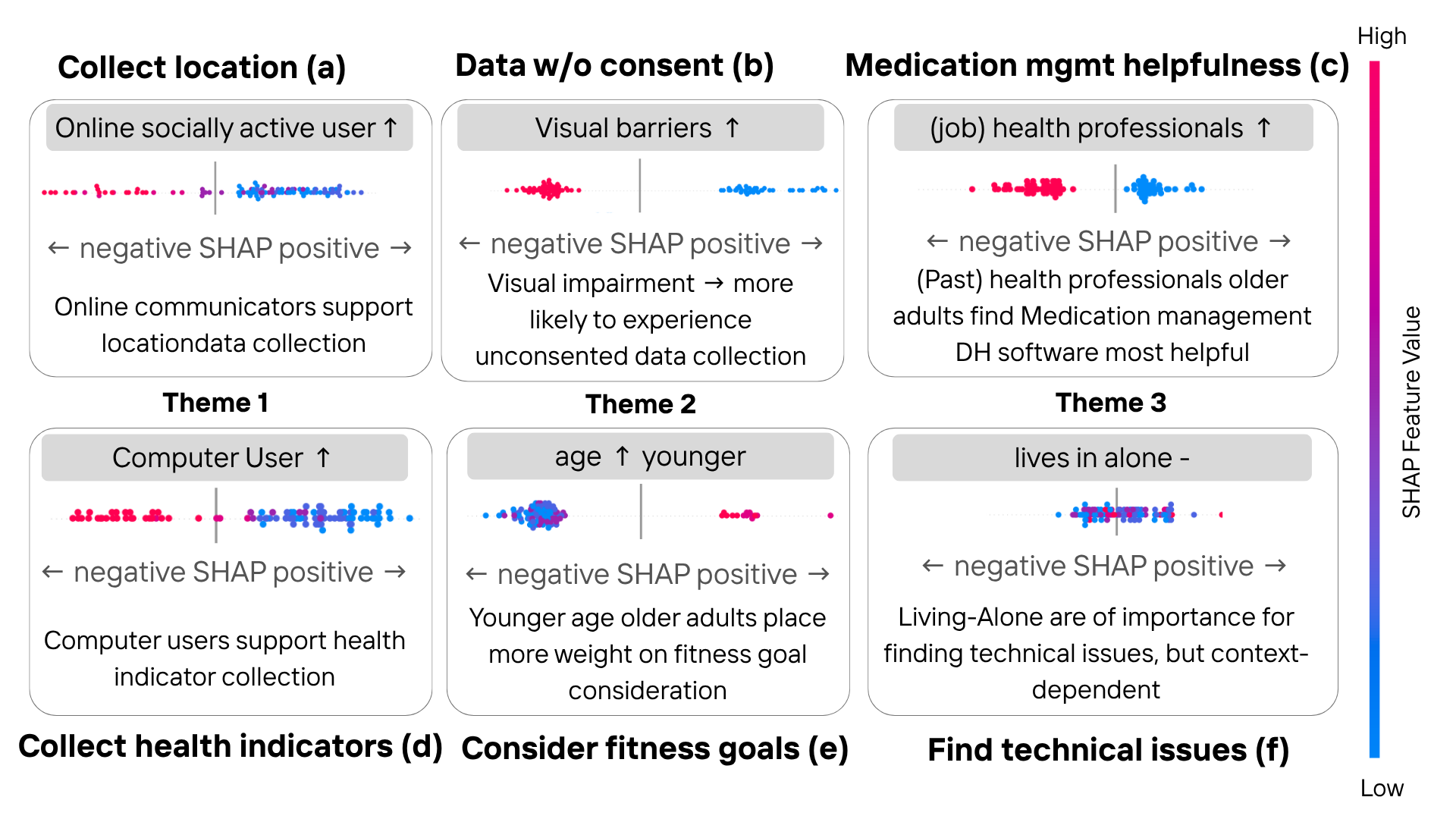}
  \caption{Key directional SHAP findings linked to interview themes}
  \label{fig:rq3_beeswarm}
\end{figure}
Figure~\ref{fig:rq3_beeswarm} presents directional SHAP findings from the OA models that were then directly validated in the interviews. Four themes emerged consistently across participants, each either confirming, refining, or contextualising a quantitative pattern.

\textbf{Theme 1: Technology access is a structural barrier shaped by device constraints, digital proficiency, and social pressure.}

Aligned with the first column results of directional findings in figure~\ref{fig:rq3_beeswarm}, this theme explains the pervasive influence of device-related and social-engagement predictors and integrates three related mechanisms. First, \textbf{device limitations}: many older adults are excluded from digital health apps because their hardware is simply inadequate. P1 observed that most elderly participants used computers already 10 years old or so'' with very low screen  resolution,'' such that they can barely see the content on the screen.'' P5 was more direct: the elderly who only have old phones simply don't have our program and can't download it.'' P4 linked this to broader socioeconomic geography, noting that digital health systems end up serving those who are really well-connected" with lower socioeconomic users disproportionately excluded. Second, \textbf{Tech proficiency} (figure~\ref{fig:rq3_beeswarm}-(a, d)): owning a capable device does not guarantee meaningful use. The SHAP distinction between computer users and social-media-active users reflects fundamentally different levels of digital embeddedness. P5 observed that smartphones are so intelligent that the elderly simply cannot learn to use them," and P8 noted that uptake depends heavily on individual awareness and existing self-care habits rather than cost alone. Third, \textbf{social and emotional pressure} (Table~\ref{tab:top5_group}): several interviewees pointed to fear of failure and the social cost of appearing incapable as barriers that are invisible to designers. P8 noted that some older adults may not accept new things due to their own cognitive decline,'' and P8 further observed that many restrict themselves because they feel they won't use such things''—a form of anticipatory self-exclusion. P11 also found that with family guidance, ``they can actually learn quickly and it's not so difficult,'' suggesting that social support mediates technology adoption more than raw capability. 

\textbf{Theme 2: Health conditions and age-related changes shape both what older adults need and what they can realistically use.}
The SHAP prominence of health conditions and the directional age effect was confirmed and contextualised by interviewees. On \textbf{visual impairment specifically} (Figure~\ref{fig:rq3_beeswarm}-(b)): P6 noted that eye problems are near-universal in this population even without formal diagnosis—it's not necessarily a disease; with age, there can also be some degenerative problems''—and that contrast sensitivity and colour differentiation (P10) are among the first practical casualties. P2 stressed that without larger text and audio fallbacks, if they really have trouble seeing, it's probably going to be quite hard to use most apps.'' Colour-coded interface conventions used by developers (P4: we give some high-contrast UI...but it can still be too hard for users who has Cataract. It can become illegible or meaningless for users with vision barriers." On \textbf{age subgroup differences}: Figure~\ref{fig:rq3_beeswarm}-(e) demonstrates that younger older adults (OA) place greater importance on fitness goals. This pattern was also supported by the interviews. P3 pointed out that adults in their 60s are still considered older adults, yet their experiences may differ substantially from those of people in their late 70s or 80s. Younger older adults may remain more fitness- and productivity-oriented, whereas older cohorts often face compounding physical and cognitive constraints---a heterogeneity that uniform ``older adult'' models inevitably flatten. P6 echoed this view, noting that ``some younger older adults who are younger than 75 can behave very differently from my other patients in their 90s; they are actually two generations, with very different exposure to technology before becoming what we generalise as `older adults'.'' This finding aligns with prior research on age-specific requirements among older adults.~\cite{mcintosh_evaluating_nodate}
\begin{findingbox}
\stepcounter{findingcount}
\noindent \textbf{Finding \arabic{findingcount}:} 
Older adults are not a uniform population—employment history, living situation, care context, and subgroup age all produce meaningfully different requirement profiles—yet these distinctions remain largely invisible in development practice, and several dimensions such as rural isolation and cognitive decline remain empirically under-characterised. Bridging this gap requires human-centric requirements engineering that accesses tech context, health context, and population heterogeneity systematically.
\end{findingbox}
\textbf{Theme 3: Older adults" is not a homogeneous group, and several under-explored human aspects remain poorly characterised.} The interviews surfaced multiple dimensions of within-group variation that the SHAP models only partially captured. On \textbf{employment and education}: Figure~\ref{fig:rq3_beeswarm}-(c) shows that \textbf{(past) health professionals} are more likely to perceive medication management as useful. This is consistent with P12, who emphasised that older adults in their 60s may still be in the workforce, including herself, and therefore place importance on ``taking medication on time'' because they are often very busy. P11 similarly reported that, as a former physician, he understands how taking half instead of a full pill can lead to a sudden drop in blood pressure, which is particularly dangerous given his moderately high blood pressure. P2, a pharmacist, also noticed that older adults with such professional experience are more likely to ``know how to use'' medication functions within Mapps. In addition, P4 noted that occupational background—particularly clinical roles such as nursing or medicine—may meaningfully influence how readily older adults interpret and engage with health data and digital health functions. This hypothesis warrants further investigation. On \textbf{living situation and care context}: the SHAP finding that \textbf{Living alone} is identified as an important predictor for technical-issue identification, although its effect was context-dependent, as the same feature value appeared with both positive and negative SHAP contributions (Figure~\ref{fig:rq3_beeswarm}-(f), Table~\ref{tab:top5_group}) may reflect a bimodal distribution. This aligns with P12's account: "Some older adults living alone are highly capable and self-managing, while others live alone — like some females in their 80s — precisely because their partner has passed away, they have no kids, and dementia makes it extra hard for them." This captures two very different profiles that produce opposing requirement priorities despite identical living-situation labels. On \textbf{under-explored aspects}: P8 identified rural location as a persistent gap, noting that digital health systems systematically exclude those without urban connectivity. Furthermore, ``Mental health conditions", ``memory impairment", and ``cognitive decline" were mentioned by multiple interviewees (P2, P3, P6, P8) as highly relevant but practically difficult to capture in survey-based studies, partly because affected individuals are rarely recruited and partly because self-report is unreliable for these dimensions. \textbf{Gender} (Table~\ref{tab:top5_group}): P9, a product manager, noted that gender may matter because most developers are male, whereas older-adult populations are statistically more likely to be female~\cite{wang2019implicit,lindau2007study}. She added, ``we did not design specifically for older women.''

\textbf{Theme 4: Stakeholders' perspectives differ systematically.}
First, \textbf{Developers systematically underestimate the
realities of older adults’ everyday technology use.}: P1 summarised the core problem: there's lots of difficulty that we never think of because we are used to good computers, we have good internet,'' whereas older adults don't even have the devices to use the software.'' P4 warned that if we make things complicated, they will not use'' the software regardless of its functional value. P6 argued for depth over breadth—software should maximise a single feature'' rather than overwhelm users—a view that directly challenges feature-maximising developer instincts. This theme explains why design-related predictors carry SHAP weight in the OA models: they capture not stylistic preference but the cumulative cost of design decisions made without adequate knowledge of users' actual environments. Second, \textbf{caregiver role type shapes requirement priorities} (also in table~\ref{tab:top5_group}). We observed clear differences between formal and informal caregivers. P2, a formal caregiver and pharmacist, highlighted a tone-related issue that mirrors the broader accessibility challenge: ``technical jargon causes users to feel alienated,'' while over-simplified graphics can make them feel ``treated like children.'' By contrast, many informal caregivers placed greater emphasis on features being ``easy to understand.'' This divergence is also reflected in themes such as \textit{current UX satisfaction} and \textit{future desired UX design} (Figure~\ref{fig:rq2_rank}). P5 highlighted this distinction by comparing her role as a formal caregiver with her role as an informal caregiver for her own older family members: ``I hope for different things for my mom than for my patients... For patients, I try my best to support recovery... But for my own parents... when they are stubborn, I can only let them decide.''

\begin{summarybox}
\textbf{Answer to RQ3:}
The interviews validated the SHAP findings while deepening their interpretation beyond what triangulation alone provides. Technology access operates through \textbf{multiple mechanisms}—device constraints, digital proficiency, and fear of failure—rather than as a single preference. Health conditions and age-cohort differences shape both what older adults value and what they can realistically use. Critically, older adults are \textbf{internally heterogeneous}: occupation, education, socioeconomic status, and rural location introduce variation that aggregate models obscure.
\end{summarybox}

% \subsection{RQ4: Implications for Automated Requirements Prioritisation}
% What to report:
% - Translate findings into an ASE contribution.
% - Propose a lightweight workflow or framework:
%   stakeholder-segmented data collection -> model training -> SHAP ranking ->
%   interview validation -> requirements prioritisation.
% - State how this supports analysts/designers in practice.
%
% Very important:
% turn findings into an actionable method, not just observations.

% \fbox{%
% \begin{minipage}{0.97\linewidth}
% \paragraph{Answer to RQ4.}
% The study suggests a stakeholder-aware automated requirements prioritisation workflow in which stakeholder segmentation, ML-based ranking, explainability analysis, and interview validation are combined into a single pipeline. This workflow enables analysts to identify which human aspects matter most, where stakeholder priorities align or diverge, and how these signals can be translated into more transparent and inclusive requirements decisions.
% \end{minipage}%
% }
\section{Discussion}
\label{sec:discussion}
% What to report:
% - Interpret the main findings.
% - Explain why stakeholder differences matter for RE and design automation.
% - Discuss what ASE researchers/practitioners learn from combining explainable ML
%   with qualitative validation.
%
% Recommended subsections:
\subsection{Stakeholder-centric Human Aspects in Aged-Care Digital Health Requirements}
Our results show that requirement priorities in aged-care digital health are not determined by demographic considerations alone; rather, they are shaped by the intersection of \textbf{health conditions}, \textbf{technology proficiency}, and \textbf{care-received context}. Across stakeholders, the most consistently important human aspects are health conditions and device usage. In particular, chronic disease, sleep disorder, and basic phone usage are strongly associated with lower satisfaction with digital health requirements. These findings further support earlier work showing that human aspects shape requirement priorities~\cite{shamsujjoha_human-centric_2021,hidellaarachchi_effects_2022, grundy_human-centric_2020}.

Across stakeholder groups, however, these influences are unevenly distributed: the most important predictors for older adults, caregivers, and developers share only a small common core (chronic disease and living situation), while the rest are group-specific. Beyond the expected patterns that caregivers emphasise patient context and health, and developers focus on functionality, one unanticipated result was that older adults prioritise technology access and care-received-related factors. This finding adds detail to prior research in software engineering and aged-care digital health that challenges the common practice of pooling multi-stakeholder input before analysis~\cite{muller_so_2022,lee_technology_2024,zainal_exploring_2025}.

For the important human aspects—\emph{technology proficiency-related}, \emph{health condition-related}, \emph{care context-related} factors, and \emph{demographics}—the interview findings deepen this interpretation. Interviewees confirmed and explained why and how these factors influence requirement priorities. On one hand, the importance of these human aspects suggests that engaging them can improve future requirements analysis and understanding. On the other, it shows that such aspects must be explicitly considered and articulated by end users and domain experts (e.g., health professionals in digital health). These findings are consistent with earlier studies~\cite{bird_generative_2021,noauthor_stakeholders_nodate}.

The practical contribution is direct. \textbf{Human-centric RE for older adults should consider not only the human aspects identified by developers, but also a broader set of human aspects by engaging end users and analysing their requirements separately.} This helps ensure that requirements prioritisation does not simply aggregate stakeholder input and encode the priorities of the most represented or articulate group while systematically obscuring others.

\subsection{Explainability as a Mechanism for Surfacing Socio-Technical Misalignment}
A second contribution of this work lies in its use of explainable machine learning with human-in-the-loop validation as an active analytical instrument. Prior work applying model-agnostic techniques such as SHAP and permutation importance in software engineering has focused on technically homogeneous tasks—defect prediction, issue classification, and effort estimation—where outcomes are well-defined and analyst populations are relatively uniform~\cite{tantithamthavorn_explainable_2022,jiarpakdee_empirical_2022}. Our study extends this logic to human-centric requirements analysis, where the goal is to identify which human aspects matter and to make visible \emph{who cares about what, and why}.

This distinction has implications for how explainable tools should support requirements engineering. Global importance rankings—the standard output of SHAP-based analyses—are informative but insufficient in multi-stakeholder settings. \textbf{Averaging importance across groups can obscure precisely the divergences that matter most for design decisions.} Instead, our results highlight the need for \emph{comparative} explainability: importance rankings decomposed by stakeholder group, enabling analysts to understand not only which aspects matter, but also to whom and under what conditions. This suggests a shift toward explainable RE toolkits that treat group-level comparison as a first-class output.

Our findings also show how human-in-the-loop validation strengthens explainability. While SHAP provides mathematical transparency, it does not guarantee interpretive clarity~\cite{jiarpakdee_empirical_2022}. Without stakeholder input, high-importance predictors may reflect causal effects, proxies, sampling artefacts, or domain-specific confounds. Our four-step integration protocol—confirming, explaining, challenging, and refining—acts as a calibration layer that distinguishes these cases and mitigates the risk of over-interpretation in explainable automation.

\subsection{Practical Guidance for Aged-Care Digital Health Teams}
Three recommendations follow from the findings. \textbf{First, target the human aspects that matter for each requirement type.} Physical health status most strongly informs functional specifications, while technical literacy drives non-functional ones. Exploiting these relationships during elicitation enables more precise requirements scoping and reduces the risk of designing for an assumed-average user who does not exist.
\textbf{Second, disaggregate stakeholder input before integrating it.} Requirements surveys and workshops should capture not only what is prioritised but under which human-aspect conditions, because the same feature carries a different meaning for an older adult managing chronic illness alone than for one supported by family. Pooling responses before this analysis does not balance perspectives; it erases the differences that matter most.
\textbf{Third, treat human aspects as first-class elicitation variables, not background metadata.} The strongest predictors in this study—health conditions, device access, living arrangement, social engagement—are routinely collected but rarely analysed as drivers of priority. We recommend capturing these variables systematically at elicitation and applying SHAP-based importance analysis to map them to specific requirement weights, establishing a traceable chain from stakeholder context through to design and validation decisions.
Taken together, these recommendations describe a form of \textbf{evidence-based, stakeholder-aware RE}: one in which explainable analytics surfaces the structure of disagreement across groups, and human judgment determines how that disagreement should be resolved.

Taken together, these recommendations point toward a form of \textbf{evidence-based, stakeholder-centric requirements engineering}: one in which explainable analytics surfaces the structure of stakeholder disagreement, and human judgment determines how that disagreement should be resolved.
\section{Implications for Automated Software Engineering}
% \label{sec:ase}
% % Optional, but useful if page budget allows.
% % What to report:
% % - Why this is an ASE paper:
% %   automation of requirements prioritisation,
% %   explainability of prioritisation decisions,
% %   stakeholder-aware decision support.
% % - Clarify the intended users of the approach:
% %   RE analysts, product teams, digital health developers.
The central contribution to ASE is a framework that makes requirements prioritisation both explainable and stakeholder-aware. Automating requirements analysis without accounting for who provided the input — and under what human-aspect conditions — does not scale inclusive RE; it scales its blind spots. The findings from RQ1–RQ3 demonstrate that explainable automation can surface the human aspects driving requirement priorities, reveal where and why stakeholder groups diverge, and provide transparent evidence to guide prioritisation decisions — turning explainability from a post-hoc audit into an active instrument for more trustworthy requirements engineering in socio-technical domains.

\begin{figure}[h]
  \centering
  \includegraphics[trim ={1cm, 11cm, 0cm, 10cm}, clip, width=\linewidth]{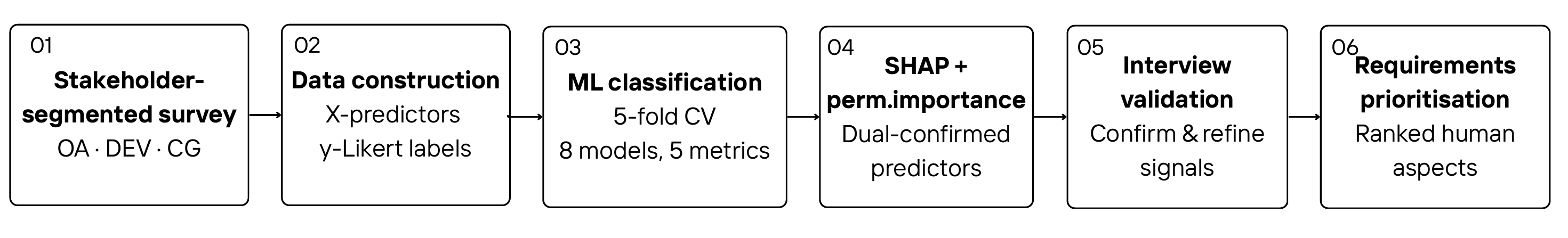}
  \caption{Proposed automated requirements prioritisation workflow}
  \label{fig:rq4_workflow}
\end{figure}

Figure~\ref{fig:rq4_workflow} summarises the lightweight workflow derived from this study.
The workflow comprises six steps: segment stakeholders into relevant groups; construct a feature matrix from human-aspect predictors and stakeholder-rated outcomes; train and evaluate ML models using cross-validation; apply SHAP and permutation importance to identify the most influential human aspects per outcome; conduct targeted interviews to confirm and refine quantitative signals; and translate confirmed patterns into ranked human aspects to support prioritisation.
This workflow contributes to ASE in two ways: it provides a repeatable pipeline combining predictive modelling with explainability in stakeholder-aware requirements analysis, and it introduces a human-in-the-loop validation stage ensuring rankings are not only statistically detectable but meaningful and actionable in practice. Rather than replacing stakeholder engagement, the workflow augments it.
In practice, analysts can use the workflow to distinguish \emph{shared} predictors — such as prior DH use, which signal broadly important requirements across groups — from \emph{group-specific} predictors — such as patient health conditions for caregivers or feature-oriented factors for developers — that surface misalignments requiring negotiation. This makes the approach particularly valuable in socio-technical contexts where requirements emerge from the interaction of end-user capability, care practice, and technical design assumptions.

\section{Threats to Validity}\label{sec:threats}This study acknowledges several limitations across construct, internal, and external validity, following the reporting standards for empirical software engineering.

\textbf{Construct Validity.} The operationalisation of human aspects'' via survey items may not fully capture the complexity of constructs like trust'' or ``socio-economics.'' A primary threat is the potential gap between \textit{stated preferences} in a survey and \textit{actual behaviour} in software usage. While surveys are standard for eliciting requirements priorities, they may not perfectly predict how stakeholders interact with a functional system. To mitigate this, we used Phase II interviews to triangulate survey responses with real-world scenarios. Furthermore, while SHAP identifies feature contributions to model predictions, these indicate model behaviour rather than causal mechanisms. We addressed this by reporting SHAP values as "associations" rather than "drivers."

\textbf{Internal Validity.} A significant threat involves \textit{selection bias} in our developer cohort. These participants were recruited via professional networks and university-affiliated labs. If the sample leans toward early-career researchers or students rather than industry practitioners, their assessment of "technical feasibility" may differ from commercial constraints. We mitigated this by collecting years of experience and role data, though the risk of a "lab-setting" perspective remains. To minimise confirmation bias in our sequential design, we employed an immersive familiarisation phase and peer debriefing, ensuring qualitative themes emerged independently of Phase I results. Additionally, as participants were not linked in real-world care dyads, interdependent dynamics (e.g., a specific caregiver’s influence on a specific patient) were not captured.

\textbf{External Validity.} Our findings reflect the socio-technical contexts of Australia and China. Online survey may have introduced a bias toward digitally literate individuals, which is a common limitation in digital health studies~\cite{xiao_requirements_2025}. Consequently, our results may over-represent the priorities of "tech-savvy" seniors. The transferability of these requirements to other clinical domains or different regulatory environments should be approached with caution. However, by including two distinct cultural contexts, we have improved the diversity of the sample compared to single-region studies.
\section{Conclusion}
\label{sec:conclusion}
This paper addressed a largely unexamined gap in requirements engineering: how human aspects differentially drive requirement priorities across stakeholder groups. Using older adults, caregivers, and developers in aged-care digital health as a representative case, we demonstrated how these differences can be identified transparently and at scale through explainable machine learning combined with qualitative validation.

Three findings stand out. First, requirement priorities are shaped by human aspects — health conditions, device access, social engagement, living context, prior digital health experience — that existing requirements methods rarely treat as first-class inputs. Second, these influences diverge sharply across groups: older adults foreground access barriers and chronic health; caregivers foreground patient context and care demands; developers foreground feature- and target-user-oriented factors, with only chronic disease and living situation emerging as shared ground. Third, this divergence is not a matter of emphasis but of kind — no single group's priorities are a reliable proxy for the others'.

The implication is direct: understanding human aspects is not a matter of adding an accessibility checklist or broadening a persona set. It means recognising that the same requirement can carry entirely different weight depending on who lives with its consequences. For digital health software serving older adults, this demands inclusive thinking at the requirements stage — accounting for the structural realities of device access, health constraints, and care context rather than assuming a capable, connected user. The workflow introduced here — ML-derived importance rankings, SHAP-based transparency, and interview-grounded validation — offers one principled path toward that goal, positioning explainability as an active mechanism for surfacing misalignment rather than a post-hoc audit. As the gap between who builds digital health systems and who depends on them continues to widen, closing it begins with requirements.
% What to report:
% - Restate problem, approach, and key findings.
% - Re-emphasize the ASE contribution:
%   explainable, stakeholder-aware automation for prioritising human aspects in requirements.
\section{Acknowledgement}
% Xiao and Grundy are supported by ARC Laureate Fellowship FL190100035. 
ChatGPT was utilised to help check the grammar in this work. 
\section{Data Availability Statement}
The anonymised replication package, including survey materials, processed data, analysis scripts, SHAP outputs, and interview protocol, is available at \texttt{10.5281/zenodo.19250920}.
\label{sec:data}
\bibliographystyle{ACM-Reference-Format}
\bibliography{sample-base}

% %%
% %% If your work has an appendix, this is the place to put it.
% \appendix









\end{document}